\documentstyle[aps,pra,twocolumn,amssymb,epsf]{revtex}
\begin{document}

\def\bea{\begin{eqnarray}}
\def\eea{\end{eqnarray}}
\def\be{\begin{equation}}
\def\ee{\end{equation}}
\def\nn{\nonumber\\}

\title{Bose-Einstein condensation of
the magnetized ideal Bose gas}

\author{Guy B. Standen\thanks{E-mail address: {\tt g.b.standen@newcastle.ac.uk}} and David J.~Toms\thanks{E-mail address:
{\tt d.j.toms@newcastle.ac.uk}}}
\address{Department of Physics, University of Newcastle Upon Tyne,\\
Newcastle Upon Tyne, United Kingdom NE1 7RU}

\date{\today}
\maketitle
\begin{abstract}
We study the charged non-relativistic Bose gas interacting with a constant magnetic field but which is otherwise free. The notion of Bose-Einstein condensation for the three dimensional case is clarified, and we show that although there is no condensation in the sense of a phase transition, there is still a maximum in the specific heat which can be used to define a critical temperature. Although the absence of a phase transition persists for all values of the magnetic field, we show how as the magnetic field is reduced the curves for the specific heat approach the free field curve. For large values of the magnetic field we show that the gas undergoes a ``dimensional reduction" and behaves effectively as a one-dimensional gas except at very high temperatures. These general features persist for other spatial dimensions $D$ and we show results for $D=5$. Finally we examine the magnetization and the Meissner-Ochsenfeld effect.
\end{abstract}
\narrowtext

The thermodynamical properties of a charged non-relativistic Bose gas in a constant magnetic field were correctly evaluated by Schafroth \cite{Schafroth55}. He showed that the presence of a constant magnetic field destroyed the familiar Bose-Einstein condensation exhibited by the free gas. Furthermore, Schafroth demonstrated that the magnetized gas exhibited the Meissner-Ochsenfeld effect which set in at a scale related to the condensation temperature of the free Bose gas. It is important to emphasize what is meant by the absence of Bose-Einstein condensation for the magnetized Bose gas. Unlike the situation for the free Bose gas, there is no phase transition (in the infinite volume limit). However this does not imply that there is no build-up of particles in the ground state. (This point has been emphasized in Ref.~\cite{Rojas,KKDJTPLB}.) Subsequent to the work of Schafroth, May~\cite{May65} looked at the magnetized Bose gas in a general spatial dimension $D$ and showed that for $D\ge5$ there was a phase transition like that in the free Bose gas. Thus the study of spatial dimensions other than 3 can be used to illustrate general features of phase transitions in magnetized systems.

We will begin with the expression for the total thermodynamic potential as found using the effective action method. (For a review see \cite{DJTKorea}.) This is
\bea
\Omega&=&\int d^Dx\left\lbrace\frac{1}{2m}|
{\mathbf D}\bar{\Psi}|^2-e\mu|\bar{\Psi}|^2\right\rbrace\nn
&&\quad+\frac{1}{\beta}\sum_n\ln\left\lbrack1-
e^{-\beta(E_n-e\mu)}\right\rbrack
\;.\label{eq1}
\eea
The first term in $\Omega$ is the ``classical" part which accounts for a possible condensate described by $\bar{\Psi}$, and the second term is the standard expression for the thermodynamic potential from statistical mechanics. Here $\beta=(kT)^{-1}$, $\mu$ is the chemical potential, and $E_n$ are the energy levels of the system. $\Omega$ in (\ref{eq1}) can be evaluated from a knowledge of the energy levels for a charged particle in a constant magnetic field \cite{LLStatMech}. Although in $D$ spatial dimensions the magnetic field may in general have more than one independent component, we will only consider the presence of a single component here. 

Whether or not there is a non-zero condensate $\bar{\Psi}\ne0$ is determined by the solution to
\be
-\frac{1}{2m}{\mathbf D}^2\bar{\Psi}-e\mu\bar{\Psi}=0\;.\label{eq5}
\ee
This is obtained by extremizing (\ref{eq1}) with respect to $\bar{\Psi}$.  It is easy to show that $\bar{\Psi}=0$ if $\mu<\mu_c$ where
$e\mu_c=E_0$
gives a critical value for the chemical potential determined by the lowest energy eigenvalue. For the case of a constant magnetic field of strength $B$ we have $E_0=\omega/2$ where $\omega=eB/m$. In this case there is no non-trivial condensate. (See Ref.~\cite{KKDJTPLB} for a general treatment.) If it is possible for $\mu$ to reach the critical value $\mu_c$, then we have $\bar{\Psi}\propto f_0({\mathbf x})$ where $f_0({\mathbf x})$ is the eigenfunction corresponding to the lowest energy.

Whether or not $\mu$ can reach $\mu_c$ is determined by the total charge $Q=-\frac{\partial\Omega}{\partial\mu}$. If we set the condensate $\bar{\Psi}=0$, so that the first term in $\Omega$ makes no contribution to the charge, and if we can solve for $\mu$ in terms of $Q$, then we must have $\mu<\mu_c$. However if it is not possible to do this then $\bar{\Psi}$ is necessarily non-zero and we have a condensate. The nature of the solutions for $\mu$ changes with temperature, and it is this behaviour which can give rise to a critical temperature signalling a phase transition. The case of a condensate ($\bar{\Psi}\ne0$) and no condensate ($\bar{\Psi}=0$) must be distinguished. From previous work \cite{May65,Tomsmag} we know that when a magnetic field is present the condition for $\bar{\Psi}\ne0$ is $D\ge5$. This contrasts with the condition for the free gas ($B=0$) which requires the spatial dimension $D\ge3$ for a non-zero condensate.

In the physically interesting case of $D=3$ the presence of a magnetic field destroys Bose-Einstein condensation if it is interpreted as a phase transition with a non-zero condensate~\cite{Schafroth55}. The presence of a magnetic field, no matter how small, results in the charged Bose gas behaving in a different manner from the same system when no magnetic field is present. However on physical grounds one might expect that for a very small magnetic field the system should behave in almost the same way as it would if $B=0$. This is one of the issues addressed in the present paper. In addition the approach of Ref.~\cite{KKDJTPLB} is clarified with a particular example.

For the free Bose gas a key feature of Bose-Einstein condensation is the behaviour of the specific heat at the critical temperature $T_0$. For $D=3$ the specific heat has a maximum at $T=T_0$, and is continuous there with a discontinuous first derivative \cite{London}. For $D=4$ the specific heat and its first derivative are continuous at $T=T_0$, and for $D\ge5$ the specific heat is discontinuous at $T=T_0$ \cite{May64}. In all cases the non-smooth behaviour of the specific heat corresponds to a maximum. As already stated the presence of a magnetic field suppresses the phase transition if $D=3,4$. For $D\ge5$ the specific heat for the magnetized gas would be expected to show a non-smooth behaviour at a critical temperature $T_c$ corresponding to the existence of a phase transition; however for $D=3,4$ the specific heat should be perfectly smooth since no phase transition occurs. We will concentrate initially on $D=3$.

By first calculating the internal energy defined by $U=\partial\Omega/(\partial\beta)$ with $\mu\beta,V,B$ held fixed, and then the specific heat at constant volume by $C_V=\partial U/(\partial T)$ with $V,B$ and the total charge $Q$ held fixed, it is possible to show that
\bea
\frac{C_V}{kN}&=&\frac{1}{\psi(1,0,1)}\bigg\lbrace x\psi(1,1,2)+\frac{3}{4}\psi(0,0,1)\nn
&&\quad+x^2(\psi(2,1,2)+2\psi(2,2,3))
-x^2\frac{\psi^2(2,1,2)}{\psi(2,0,1)}\bigg\rbrace\nn
&&\quad-\frac{1}{4}\frac{\psi(1,0,1)}{\psi(2,0,1)}
-x\frac{\psi(2,1,2)}{\psi(2,0,1)}\;,\label{eq9}
\eea
where 
\be
N=\frac{Q}{e}=V\left(\frac{m}{2\pi\beta}\right)^{3/2}x\psi(1,0,1)
\;,\label{eq10}
\ee
is the total number of particles present. Here $\psi(a,b,c)$ denotes a class of sums defined by
\be
\psi(a,b,c)=\sum_{l=1}^{\infty}l^{a-D/2}e^{-l(\epsilon+b)x}(1-e^{-lx})^{-c}\;.\label{eq11}
\ee
(We take $D=3$ in eqs.(\ref{eq9}) and (\ref{eq10}).)

\begin{figure}[htb]
\epsfxsize=252pt
\epsfbox{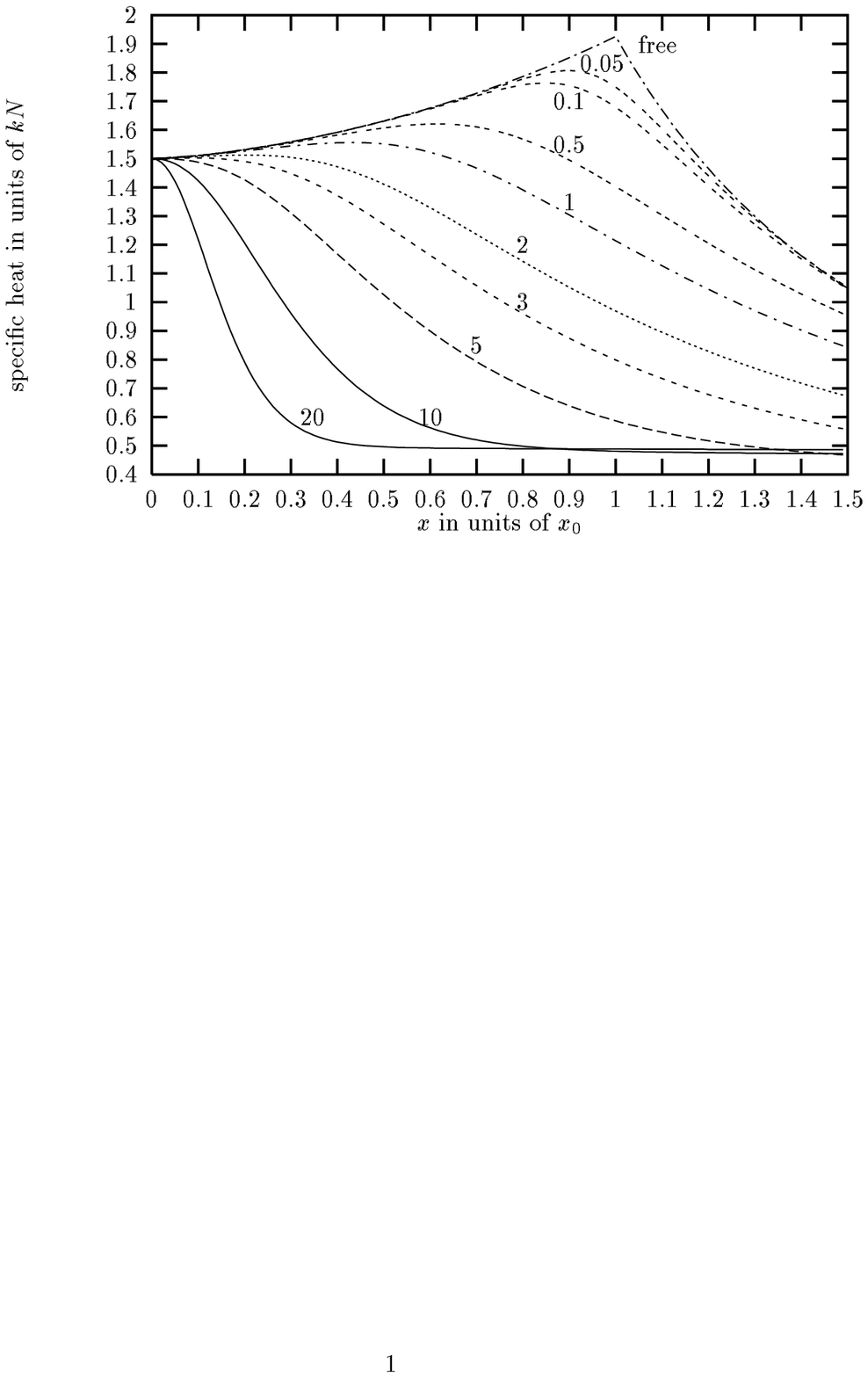}
\caption{\narrowtext This shows the specific heat for the magnetized Bose gas in units of $kN$. The curve labelled free is the result for the uncharged Bose gas. The $x$ coordinate in the figure is $x=\hbar\omega/(kT)$ in units of $x_0=\hbar\omega/(kT_0)$ with $T_0$ the critical temperature for the free Bose gas. The numbers labelling the curves are the values for $x_0$. The smaller numbers indicate lower values for the magnetic field.\label{fig1}}
\end{figure}

It is possible to solve (\ref{eq10}) for $\epsilon$ where $\mu=\omega(1/2-\epsilon)$ and then to calculate $C_V$. The result of this is shown in fig.~\ref{fig1}. For comparison we have also indicated the result for the free Bose gas. As expected the specific heat is perfectly continuous for $B\ne0$. As $B$ is decreased (following the curves with smaller values of $x_0$) although $C_V$ remains continuous the peak starts to increase and the curve starts to resemble that for the free Bose gas. The maximum in the specific heat becomes sharper as $B$ is decreased with the temperature of the maximum approaching the temperature $T_0$ for the free Bose gas. Similar results hold if the number of particles in the ground state is computed. Thus although for $B\ne0$ the system does not have a phase transition characterized by a critical temperature and a non-zero condensate $\bar{\Psi}$ and is therefore very different from the same system with $B=0$, quantities of direct physical interest, such as the specific heat or ground state particle number, are closely approximated by the free Bose gas result for small enough values of $B$. We will obtain approximate analytical expressions for small $B$ later.

It is also of interest to observe what happens as the value of the magnetic field is increased. It is clear from the curves in fig.~\ref{fig1} with larger values of $x_0$ that with increasing $B$ the peak in $C_V$ shifts to smaller values of $x$ (which corresponds to higher temperatures) and decreases in magnitude. (Although the resolution of the graphs do not show it, there is always a maximum in the specific heat.) As $x$ approaches 0, corresponding to infinitely high temperatures, the specific heat approaches the classical Maxwell-Boltzmann result of 1.5 regardless of the magnitude of $B$, as would be expected. For large $B$ the specific heat actually resembles that for a free Bose gas in one spatial dimension. This can be understood heuristically by the fact that classically the orbit of a charged particle in a constant magnetic field is a spiral around the direction specified by the magnetic field. The size of the orbit is inversely proportional to the magnetic field; thus as $B$ is increased the motion of the particle becomes more and more one-dimensional. This substantiates the general point of view used in Ref.~\cite{KKDJTPLB} in which the leading behaviour of thermodynamic quantities is well approximated by studying the lowest energy solution. In this case as $B$ is increased the gap between the ground and first excited state becomes increasingly large and the ground state makes the leading contribution.

Although for $B\ne0$ with $D=3$ there is no phase transition, Schafroth~\cite{Schafroth55} showed that the magnetized gas exhibited the Meissner-Ochsenfeld effect. He found
\be
M\simeq-\frac{\rho}{2m}\left\lbrack1-\left(\frac{x_0}{x}\right)^{3/2}\right\rbrack\label{eq12}
\ee
where $\rho$ is the charge density and $x_0=\omega/(kT_0)$. The approximation assumes that $T\le T_0$. Using $M=\partial\Omega/(\partial B)$, the exact expression may be computed to be
\be
M=-\frac{\rho}{2m}\left\lbrace1-
\frac{2\psi(0,0,1)-2x\psi(1,1,2)}{x\psi(1,0,1)}\right\rbrace\;.\label{eq13}
\ee
This expression holds for any temperature.
\begin{figure}[htb]
\epsfxsize=252pt
\epsfbox{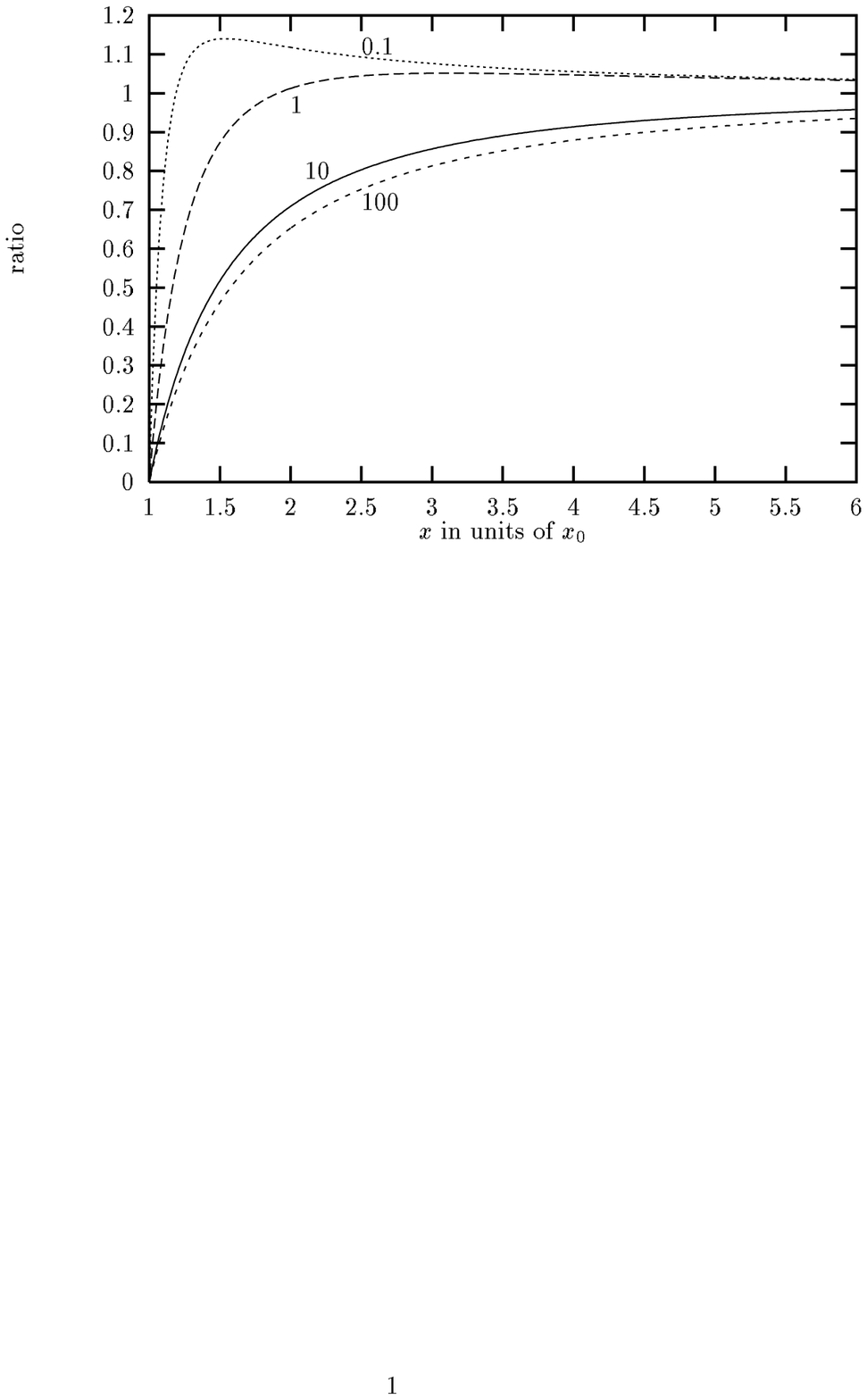}
\caption{\narrowtext This shows the ratio of the magnetization computed using the Schafroth approximation in eq.~(\ref{eq12}) to the exact result. The labels 100,10,1,0.1 on the curves indicate the value of $x_0$, the smaller numbers corresponding to lower values for the magnetic field. \label{fig2}}
\end{figure}
The ratio of the Schafroth approximation to the true result is plotted in fig.~\ref{fig2}. As $B$ is lowered the Schafroth approximation becomes better over a wider range of temperature, and as $T\rightarrow T_0$. As noted by Schafroth~\cite{Schafroth55}, the approximation will break down when $T$ becomes too close to $T_0$, and this is borne out by the detailed calculation. For $T\simeq T_0/10$ the agreement is within 2\%.

It is possible to derive the Schafroth approximation in (\ref{eq12}) from the exact result (\ref{eq13}) by approximating for the class of sums (\ref{eq11}). A powerful technique for doing this makes use of the Mellin-Barnes integral representation for the exponential function \cite{DFPRD,KKDJTPLB}. In Ref.~\cite{DFPRD} it was shown that for $\epsilon\simeq 1/2$ (corresponding to $\mu\simeq0$) there was a superdiamagnetic regime where the magnetization $M\propto -B^{1/2}$. (A previous paper which studied the magnetization is Ref.~\cite{Arias}, although the superdiamagnetism was not found.) The temperature range at which $\epsilon\simeq1/2$ is close to but slightly above $T_0$. We have obtained results in a temperature range which includes $T=T_0$. By solving (\ref{eq10}) for $\epsilon$ we find
\bea
\epsilon&\simeq& a+\frac{2\zeta(1/2)}{\pi^{1/2}\zeta_H(3/2,a)}
\left(\frac{1}{2}-a\right)x_0^{1/2}\nn
&&\quad+\frac{6\zeta(3/2)}{\pi^{1/2}\zeta_H(3/2,a)x_0^{1/2}}
\left\lbrack1-\left(\frac{x}{x_0}\right)^{1/2}\right\rbrack\;,
\label{eq14}
\eea
which is good if $1-(x/x_0)^{1/2}<<x_0^{1/2}$. Here $a$ is just a constant defined by $\zeta_H(1/2,a)=0$ with $\zeta_H$ the Hurwitz $\zeta$-function. (Numerically we find $a\simeq0.302721829$.) As $x_0$ is decreased the range over which the expansion holds becomes increasingly small, but it is always good at $x=x_0$ (corresponding to $T=T_0$). The agreement between the approximation (\ref{eq14}) and a numerical evaluation of $\epsilon$ is very good for small $x_0$.
\begin{figure}[b]
\epsfxsize=252pt
\epsfbox{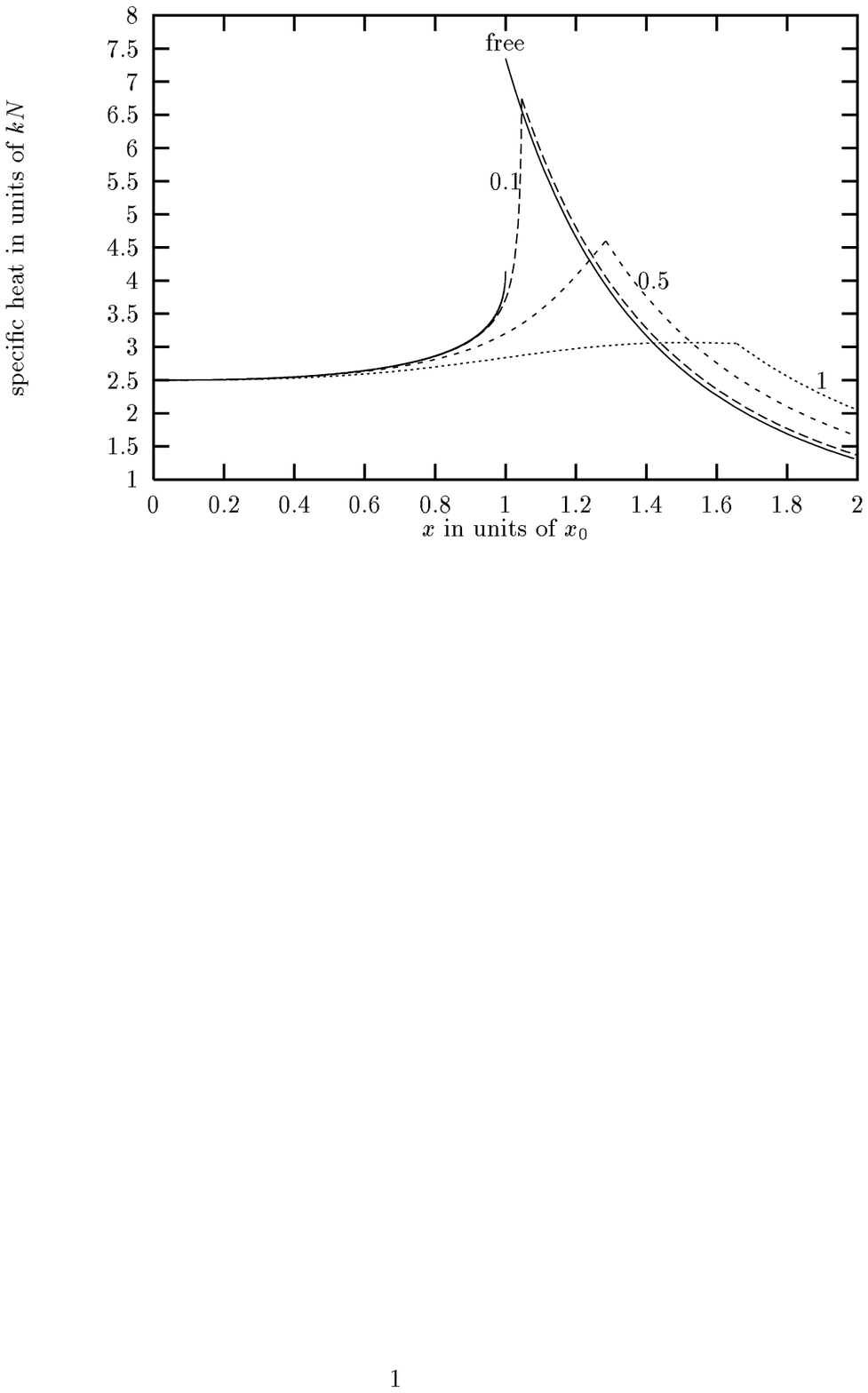}
\caption{\narrowtext This shows the specific heat for small values of the magnetic field. The labels on the curve indicate the results for $x_0=0.1,0.5,1$. For comparison the result for no magnetic field is shown in the discontinuous curve. \label{fig3}}
\end{figure}
Although it might be thought that there was very little difference between our result of $\epsilon\simeq0.3$ and the result $\epsilon\simeq0.5$
of Ref.~\cite{DFPRD}, in fact the magnetization and other thermodynamic quantities are extremely sensitive to the value of $\epsilon$. Using our expansion (\ref{eq14}) we find
\bea
M&\simeq&-\frac{\rho}{2m}\Big\lbrace
\frac{6\pi^{1/2}\zeta_H(-1/2,a)}{\zeta(3/2)}x_0^{1/2}\nn
&&\quad-2\frac{\zeta(1/2)}{\zeta(3/2)}(\frac{1}{6}-a+a^2)x_0\Big\rbrace
\;,\label{eq15}
\eea
for small $x_0$. This shows the $-B^{1/2}$ behaviour found in Ref.~\cite{DFPRD}, but with a slightly different numerical factor due to the lower temperature range we are using. 

In addition to the magnetization we have obtained an approximate expression for the specific heat using (\ref{eq14}) which is
\be
\frac{C_V}{kN}\simeq\frac{15}{4}\frac{\zeta(5/2)}{\zeta(3/2)}-
\frac{9\zeta(3/2)}{2\pi^{1/2}\zeta_H(3/2,a)}x_0^{1/2}
\;,\label{eq16}
\ee
at $T=T_0$. The first term in (\ref{eq16}) which survives in the $B\rightarrow0$ limit may be observed \cite{London} to be the specific heat for the free Bose gas at the critical temperature $T_0$. The presence of a non-zero magnetic field is seen to lower the specific heat from the free field result which is consistent with the results shown in fig.~1. It is possible to systematically determine higher order terms in the expansions (\ref{eq14}--\ref{eq16}) and increase the accuracy of the approximation. This and other details will be presented elsewhere \cite{GBSDJT}.

As an example of a magnetic system where a phase transition does occur we will consider the case $D=5$. When $B=0$ the specific heat is discontinuous at the critical temperature $T_0$. For $B\ne0$ the specific heat is given by an expression similar to (\ref{eq9}), and is perfectly continuous. The results are shown in fig.~\ref{fig3}.

\begin{figure}[htb]
\epsfxsize=252pt
\epsfbox{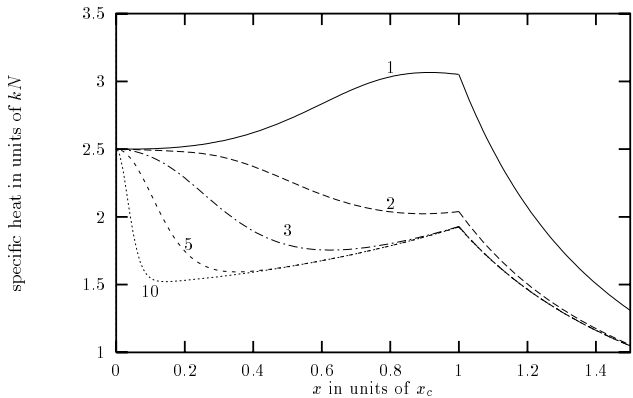}
\caption{\narrowtext This shows the specific heat for small values of the magnetic field. The labels on the curve indicate the results for $x_0=1,2,3,5,10$.  \label{fig4}}
\end{figure}

The critical temperature $T_c$ for the phase transition cannot be computed analytically if $B\ne0$; however for small values of the magnetic field it is possible to show that \cite{DJTKorea}
\be
kT_c\simeq kT_0-\frac{\zeta(3/2)}{5\zeta(5/2)}\frac{eB}{m}\;.
\label{eq17}
\ee
The critical temperature is lowered from that for the free Bose gas when there is a non-zero magnetic field. The approximation in (\ref{eq17}) is in good agreement with the exact result for small $B$ but breaks down as $B$ becomes large \cite{GBSDJT}. The results in fig.~\ref{fig3} show that as $B$ is decreased the curve for the specific heat starts to resemble the discontinuous one for the free gas. The critical temperature approaches the free field result, which is clear from the small $B$ approximation in (\ref{eq17}). The height of the specific heat maximum is seen to increase in magnitude and the slope of the curve for $T>T_c$ becomes steeper as $B$ is reduced. On the other hand as the magnetic field is increased in magnitude the specific heat undergoes a marked change. The results for larger values of the magnetic field are shown in fig.~\ref{fig4}. 
For very large values of $B$ the curves for the specific heat resemble the result found for the free Bose gas in three spatial dimensions. This provides a further and striking example of the ``dimensional reduction" in the presence of a magnetic field.

In conclusion, we have presented the results of a study of the ideal charged Bose gas in a constant magnetic field. For certain values of the temperature and magnetic field we have obtained analytical results and compared them with numerical ones. By examining the specific heat it was shown how in a strong magnetic field the system exhibits a ``dimensional reduction" where the system behaves as if the spatial dimension was lowered by 2. We have also examined the way in which the free Bose gas results are approached for small $B$. Although there is a significant difference between the case $B=0$ (which has a phase transition) and the $B\ne0$ gas (which has no phase transition), for small values of $B$ there is very little difference between the two specific heat curves. Finally for spatial dimensions $D\ge5$ a proper account of the condensate is essential for obtaining correct results, and this can be done in an efficient manner by using the effective action method.

GBS is grateful to the EPSRC for grant 94004194. DJT would like to thank K. Kirsten, F. Laloe, and S. Ouvry for helpful discussions. We are both grateful to J. Daicic and N. Frankel for helping us to clarify comments concerning Refs.~\cite{DFPRD,Arias}.

\end{document}